\documentclass[sigconf,table]{acmart} %

\AtBeginDocument{%
  \providecommand\BibTeX{{%
    \normalfont B\kern-0.5em{\scshape i\kern-0.25em b}\kern-0.8em\TeX}}}

\copyrightyear{2021} 
\acmYear{2021} 
\setcopyright{rightsretained} 
\acmConference[FAccT '21]{Conference on Fairness, Accountability, and Transparency}{March 3--10, 2021}{Virtual Event, Canada}
\acmBooktitle{Conference on Fairness, Accountability, and Transparency (FAccT '21), March 3--10, 2021, Virtual Event, Canada}\acmDOI{10.1145/3442188.3445932}
\acmISBN{978-1-4503-8309-7/21/03}

\usepackage{xcolor} 
\usepackage{amsmath}
\usepackage{amsfonts}
\usepackage{wrapfig}
\usepackage{sidecap}
\usepackage{subcaption}
\usepackage{array,multirow,textcomp}
\usepackage{colortbl}
\usepackage{enumitem}
\definecolor{d_small}{rgb}{0.90,0.90,1}
\definecolor{d_medium}{rgb}{0.80,0.80,1}
\definecolor{d_large}{rgb}{0.70,0.70,1}
\definecolor{gray}{rgb}{0.9,0.9,0.9}

\newif\ifblind
\newif\ifdraft
\newif\ifarxiv

\newcommand{\github}{\ifblind\href{https://github.com/anonymous/repo}{github.com/anonymous/repo}\else \href{https://github.com/ryansteed/ieat}{github.com/ryansteed/ieat}\fi}

\newcommand{\ttitle}{Image Representations Learned With Unsupervised Pre-Training Contain Human-like Biases}

\arxivfalse %
\blindfalse
\draftfalse

\begin{document}

\title{\ttitle}

\author{Ryan Steed}
\email{ryansteed@cmu.edu}
\affiliation{%
  \institution{Carnegie Mellon University}
  \city{Pittsburgh}
  \state{Pennsylvania}
  \country{USA}
}

\author{Aylin Caliskan}
\email{aylin@gwu.edu}
\affiliation{%
  \institution{George Washington University}
  \city{Washington}
  \state{District of Columbia}
  \country{USA}
}

\begin{abstract}
Recent advances in machine learning leverage massive datasets of unlabeled images from the web to learn general-purpose image representations for tasks from image classification to face recognition. But do unsupervised computer vision models automatically learn implicit patterns and embed social biases that could have harmful downstream effects? We develop a novel method for quantifying biased associations between representations of social concepts and attributes in images. We find that state-of-the-art unsupervised models trained on ImageNet, a popular benchmark image dataset curated from internet images, automatically learn racial, gender, and intersectional biases. We replicate 8 previously documented human biases from social psychology, from the innocuous, as with insects and flowers, to the potentially harmful, as with race and gender. Our results closely match three hypotheses about intersectional bias from social psychology. For the first time in unsupervised computer vision, we also quantify implicit human biases about weight, disabilities, and several ethnicities.
 When compared with statistical patterns in online image datasets, our findings suggest that machine learning models can automatically learn bias from the way people are stereotypically portrayed on the web.
 
\end{abstract}

\ifarxiv\else

\fi

\maketitle

\section{Introduction}
\label{sec:introduction}
Can machines learn social biases from the way people are portrayed in image datasets? Companies and researchers regularly use machine learning models trained on massive datasets of images scraped from the web for tasks from face recognition \citep{Hill2020TheIt} to image classification \citep{Sun2017RevisitingEra}. To reduce costs, many practitioners use state-of-the-art models ``pre-trained" on large datasets to help solve other machine learning tasks, a powerful approach called \emph{transfer learning} \citep{Tan2018ALearning}. For example, HireVue used similar state-of-the-art computer vision and natural language models to evaluate job candidates' video interviews, potentially discriminating against candidates based on race, gender, or other social factors \citep{Harwell2019AJob}. In this paper, we show how models trained on unlabeled images scraped from the web embed human-like biases, including racism and sexism.

Where most bias studies focus on supervised machine learning models, we seek to quantify learned patterns of implicit social bias in unsupervised image representations. Studies in supervised computer vision have highlighted social biases related to race, gender, ethnicity, sexuality, and other identities in tasks including face recognition, object detection, image search, and visual question answering \citep{Buolamwini2018GenderClassification, Kay2015UnequalOccupations, raji2020saving, Wilson2019PredictiveDetection, Manjunatha2019ExplicitModels, Nex2014UAVReview}. These algorithms are used in important real-world settings, from applicant video screening \citep{Harwell2019AJob, Raghavan2020MitigatingPractices} to autonomous vehicles \citep{Geiger2012AreSuite, Nex2014UAVReview}, but their harmful downstream effects have been documented in applications such as online ad delivery \citep{Sweeney1997Weaving1997} and image captioning \citep{Hendricks2018WomenModels}.

\begin{figure}[!t]
    \begin{center}
        \includegraphics[width=0.32\textwidth]{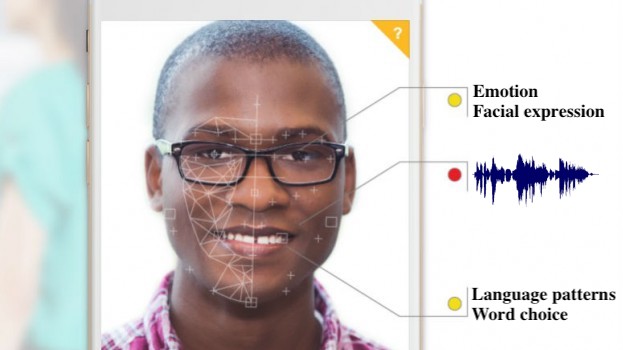}
        \label{fig:hirevure}
    \end{center}
    \caption{Unilever using AI-powered job candidate assessment tool HireVue \citep{Harwell2019AJob}.}
\end{figure}

Our work examines the growing set of computer vision methods in which no labels are used during model training. Recently, pre-training approaches adapted from language models have dramatically increased the quality of unsupervised image representations \citep{Donahue2019LargeLearning, Bachman2019LearningViews, He2020MomentumLearning, Chen2020ARepresentations, Chen2020DeepIntervals, Chen2020GenerativePixels, Misra2020Self-SupervisedRepresentations, Carion2020End-to-EndTransformers}. With \emph{fine-tuning}, practitioners can pair these general-purpose representations with labels from their domain to accomplish a variety of supervised tasks like face recognition or image captioning. We hypothesize that 1) like their counterparts in language, these unsupervised image representations also contain human-like social biases, and 2) these biases correspond to stereotypical portrayals of social group members in training images.

Results from natural language support this hypothesis. Several studies show that word embeddings, or representations, learned automatically from the way words co-occur in large text corpora exhibit human-like biases \citep{Bolukbasi2016ManEmbeddings,Caliskan2017,Garg2018WordStereotypes}. Word embeddings acquire these biases via statistical regularities in language that are based on the co-occurrence of stereotypical words with social group signals. Recently, new deep learning methods for learning context-specific representations sharply advanced the state-of-the-art in natural language processing (NLP) \citep{Devlin2018Bert:Understanding, Peters2018DeepRepresentations, Radford2019LanguageLearners}. Embeddings from these pre-trained models can be fine-tuned to boost performance in downstream tasks such as translation \citep{Erhan2009ThePre-training, Erhan2010WhyLearning}. As with static embeddings, researchers have shown that embeddings extracted from contextualized language models also exhibit downstream racial and gender biases \citep{Zhao2017MenConstraints, Basta2019EvaluatingEmbeddings, Tan2019AssessingRepresentations, Guo2020DetectingBiases}. 

Recent advances in NLP architectures have inspired similar unsupervised computer vision models. We focus on two state-of-the-art, pre-trained models for image representation, iGPT \citep{Chen2020GenerativePixels} and SimCLRv2 \citep{Chen2020DeepIntervals}. We chose these models because they hold the highest fine-tuned classification scores, were pre-trained on the same large dataset of Internet images, and are publicly available. iGPT, or Image GPT, borrows its architecture from GPT-2 \citep{Radford2019LanguageLearners}, a state-of-the-art unsupervised language model. iGPT learns representations for pixels (rather than for words) by pre-training on many unlabeled images \citep{Chen2020GenerativePixels}. 
SimCLRv2 uses deep learning to construct image representations from ImageNet by comparing augmented versions of the training images \citep{Chen2020ARepresentations, Chen2020DeepIntervals}.

Do these unsupervised computer vision models embed human biases like their counterparts in natural language? If so, what are the origins of this bias? In NLP, embedding biases have been traced to word co-occurrences and other statistical patterns in text corpora used for training \citep{Caliskan2017, Brunet2019UnderstandingEmbeddings, Blodgett2020LanguageNLP}. Both our models are pre-trained on ImageNet 2012, the most widely-used dataset of curated images scraped from the web \citep{Russakovsky2015ImageNetChallenge}.  In image datasets and image search results, researchers have documented clear correlations between the presence of individuals of a certain gender and the presence of stereotypical objects; for instance, the category ``male" co-occurs with career and office related content such as ties and suits whereas ``female" more often co-occurs with flowers in casual settings \citep{Kay2015UnequalOccupations, Wang2020REVISE:Datasets}. As in NLP, we expect that these patterns of bias in the pre-training dataset will result in implicitly embedded bias in unsupervised models, even without access to labels during training.

This paper presents the Image Embedding Association Test (iEAT), the first systematic method for detecting and quantifying social bias learned automatically from unlabeled images.

\begin{itemize}[leftmargin=*]
    \item We find statistically significant racial, gender, and intersectional biases embedded in two state-of-the-art unsupervised image models pre-trained on ImageNet \citep{Russakovsky2015ImageNetChallenge}, iGPT \citep{Chen2020GenerativePixels} and SimCLRv2 \citep{Chen2020DeepIntervals}.
    \item We test for 15 previously documented human and machine biases that have been studied for decades and validated in social psychology and conduct the first machine replication of Implicit Association Tests (IATs) with picture stimuli \citep{Greenwald1998MeasuringTest}. 
    \item In 8 tests, our machine results match documented human biases, including 4 of 5 biases also found in large language models. The 7 tests which did not show significant human-like biases are from IATs with only small samples of picture stimuli.
    \item With 16 novel tests, we show how embeddings from our model confirm several hypotheses about intersectional bias from social psychology \citep{Ghavami2013AnHypotheses}.
    \item We compare our results to statistical analyses of race and gender in image datasets. Unsupervised models seem to learn bias from the ways people are commonly portrayed in images on the web.
    \item We present a qualitative case study of how image generation, a downstream task utilizing unsupervised representations, exhibits a bias towards the sexualization of women.
\end{itemize}

\section{Related Work}
\label{sec:related}

Various tests have been constructed to quantify bias in unsupervised natural language models \citep{Caliskan2017, Zhao2017MenConstraints, Basta2019EvaluatingEmbeddings, May2019OnEncoders}, but to our knowledge, there are no principled tests for measuring bias embedded in \textit{unsupervised} computer vision models. \citet{Wang2020REVISE:Datasets} develop a method to automatically recognize bias in visual datasets but still rely on human annotations. Our method uses no annotations whatsoever.
In NLP, there are several systematic approaches to measuring unsupervised bias in word embeddings \citep{Caliskan2017, May2019OnEncoders, Tan2019AssessingRepresentations, Guo2020DetectingBiases, Bommasani2020InterpretingEmbeddings, Kurita2019MeasuringRepresentations}. Most of these tests take inspiration from the well-known IAT \citep{Greenwald1998MeasuringTest, Greenwald2003UnderstandingAlgorithm}. Participants in the IAT are asked to rapidly associate stimuli, or exemplars, representing two target concepts (e.g. ``flowers" and ``insects") with stimuli representing evaluative attributes (e.g. ``pleasant" and ``unpleasant") attribute \citep{Greenwald1998MeasuringTest}. Assuming that the cognitive association task is easier when the strength of implicit association between the target concept and attributes is high, the IAT quantifies bias as the latency of response \citep{Greenwald1998MeasuringTest} or the rate of classification error \citep{Nosek2001TheTask}. Stimuli may take the form of words, pictures, or even sounds \citep{Nosek2007TheReview}, and there are several IATs with picture-only stimuli \citep{Nosek2007TheReview}.

Notably, \citet{Caliskan2017} adapt the heavily-validated IAT \citep{Greenwald1998MeasuringTest} from social psychology to machines by testing for the mathematical association of word embeddings rather than response latency. They present a systematic method for measuring language biases associated with social groups, the Word Embedding Association Test (WEAT). Like the IAT, the WEAT measures the effect size of bias in static word embeddings by quantifying the relative associations of two sets of target stimuli (e.g., \{``woman," ``female"\} and \{``man," ``male"\}) that represent social groups with two sets of evaluative attributes (e.g., \{``science," mathematics"\} and \{``arts," ``literature"\}). For validation, two WEATs quantify associations towards flowers vs. insects and towards musical instruments vs. weapons, both accepted baselines \citet{Greenwald1998MeasuringTest}. \citet{Greenwald1998MeasuringTest} refer to these baseline biases as ``universally" accepted stereotypes since they are widely shared across human subjects and are not potentially harmful to society. Other WEATs measure social group biases such as sexist and racist associations or negative attitudes towards the elderly or people with disabilities. In any modality, implicit biases can potentially be prejudiced and harmful to society. If downstream applications use these representations to make consequential decisions about human beings, such as automated video job interview evaluations, machine learning may perpetuate existing biases and exacerbate historical injustices \cite{Raghavan2020MitigatingPractices, De-Arteaga2019BiasSetting}.

The original WEAT \citep{Caliskan2017} uses \emph{static} word embedding models such as word2vec \citep{Mikolov2013EfficientSpace} and GloVe \cite{Pennington2014Glove:Representation}, each trained on Internet-scale corpora composed of billions of tokens. Recent work extends the WEAT to \emph{contextualized} embeddings: dynamic representations based on the context in which a token appears. \citet{May2019OnEncoders} insert targets and attributes into sentences like ``This is a[n] <word>" and applying WEAT to the vector representation for the whole sentence, with the assumption that the sentence template used is ``semantically bleached" (such that the only meaningful content in the sentence is the inserted word). \citet{Tan2019AssessingRepresentations} extract the contextual word representation for the token of interest before pooling to avoid confounding effects at the sentence level; in contrast, \citet{Bommasani2020InterpretingEmbeddings} find that pooling tends to improve representational quality for bias evaluation. \citet{Guo2020DetectingBiases} dispense with sentence templates entirely, pooling across $n$ word-level contextual embeddings for the same token extracted from random sentences. Our approach is closest to these latter two methods, though we pool over images rather than words.

\section{Approach}
\label{sec:approach}
In this paper, we adapt bias tests designed for contextualized word embeddings to the image domain. While language transformers produce contextualized \emph{word} representations to solve the next \emph{token} prediction task, an image transformer model like iGPT generates \emph{image} representations to solve the next \emph{pixel} prediction task  \citep{Chen2020GenerativePixels}. Unlike words and tokens, pixels do not explicitly correspond to semantic concepts (objects or categories) as words do.  In language, a single token (e.g. ``love") corresponds to the target concept or attribute (e.g. ``pleasant"). But in images, no single pixel corresponds to a semantically meaningful concept. To address the abstraction of semantic representation in the image domain, we propose the Image Embedding Association Test (iEAT), which modifies contextualized word embedding tests to compare pooled image-level embeddings. The goal of the iEAT is to measure the biases embedded during unsupervised pre-training by comparing the relative association of image embeddings in a systematic process. \citet{Chen2020GenerativePixels} and \citet{Chen2020ARepresentations} show through image classification that unsupervised image features are good representations of object appearance and categories; we expect they will also embed information gleaned from the common co-occurrence of certain objects and people and therefore contain related social biases.

\begin{figure}[!t]
    \centering
    \includegraphics[width=1\linewidth]{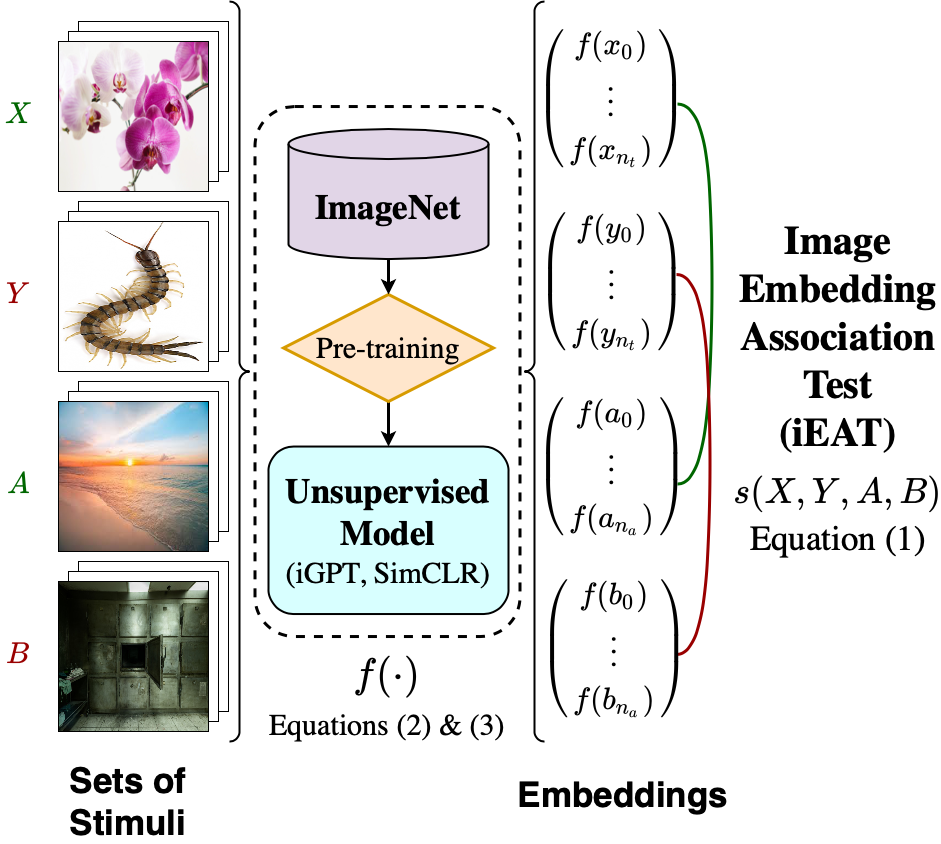}
    \caption{Example iEAT replication of the Insect-Flower IAT \citep{Greenwald1998MeasuringTest}, which measures the differential association between flowers vs. insects and pleasantness vs. unpleasantness.}
    \label{fig:diagram}
\end{figure}

Our approach is summarized in Figure~\ref{fig:diagram}. The iEAT uses the same formulas for the test statistic, effect size $d$, and $p$-value as the WEAT \citep{Caliskan2017}, described in Section~\ref{subsec:eats}. Section~\ref{subsec:replications} summarizes our approach to replicating several different IATs; Section~\ref{subsec:intersectional} describes several novel intersectional iEATs. Section~\ref{subsec:eats} describes our test statistic, drawn from embedding association tests like the WEAT.

\subsection{Replication of Bias Tests}
\label{subsec:replications}
In this paper, we validate the iEAT by replicating as closely as possible several common IATs. These tests fall into two broad categories: valence tests, in which two target concepts are tested for association with ``pleasant" and ``unpleasant" images; and stereotype tests, in which two target concepts are tested for association with a pair of stereotypical attributes (e.g. ``male" vs. ``female" ``career" vs. ``family"). To closely match the ground-truth human IAT data and validate our method, our replications use the same concepts as the original IATs (listed in Table~\ref{tab:results}). Because some IATs rely on verbal stimuli, we adapt them to images, using image stimuli from the IATs when available. When no previous studies use image stimuli, we map the non-verbal stimuli to images using the data collection method described in Section~\ref{sec:data}.

Many of these bias tests have been replicated for machines in the language domain; for the first time, we also replicate tests with image-only stimuli, including the Asian and Native American IATs. Most of these tests were originally administered in controlled laboratory settings \citep{Greenwald1998MeasuringTest, Greenwald2003UnderstandingAlgorithm}, and all except for the Insect-Flower IAT have also been tested on the Project Implicit website at \url{http://projectimplicit.org} \citep{Nosek2002HarvestingSite, Greenwald2003UnderstandingAlgorithm, Greenwald2009UnderstandingValidity.}. Project Implicit has been available worldwide for over 20 years; in 2007, the site had collected more than 2.5 million IATs. The average effect sizes (which are based on samples so large the power is nearly 100\%) for these tests are reproduced in Table~\ref{tab:results}. To establish a principled methodology, all the IAT verbal and original image stimuli for our bias tests were replicated exactly from this online IAT platform \citep{Nosek2007PervasivenessStereotypes}. We will treat these results, along with the laboratory results from the original experiments \citep{Greenwald1998MeasuringTest}, as ground-truth for human biases that serve as validation benchmarks for our methods (Section~\ref{sec:evaluation}).

\subsection{Intersectional iEATs}
\label{subsec:intersectional}
We also introduce several new tests for intersectional valence bias and bias at the intersection of gender stereotypes and race. Intersectional stereotypes are often even more severe than their constituent stereotypes \citep{Crenshaw1990MappingColor}. Following \citet{Tan2019AssessingRepresentations}, we anchored comparison on White males, the group with the most representation, and compared against White females, Black males, and Black females, respectively (Table~\ref{tab:intersectional}). Drawing on social psychology \citep{Ghavami2013AnHypotheses}, we pose three hypotheses about intersectional bias: 
\begin{itemize}[leftmargin=*]
    \item \emph{Intersectionality hypothesis}: tests at the intersection of gender and race will reveal emergent biases not explained by the sum of biases towards race and gender alone.
    \item \emph{Race hypothesis}: biases between racial groups will be more similar to differential biases between the men than between the women.
    \item \emph{Gender hypothesis}: biases between men and women will be most similar to biases between White men and White women.
\end{itemize}

\subsection{Embedding Association Tests}
\label{subsec:eats}
Though our stimuli are images rather than words, we can use the same statistical method for measuring biased associations between image representations \citep{Caliskan2017} to quantify a standardized effect size of bias. We follow \citet{Caliskan2017} in describing the WEAT here. 

Let $X$ and $Y$ be two sets of target concepts embeddings of size $N_t$, and let $A$ and $B$ be two sets of attribute embeddings of size $N_a$. For example, the Gender-Career IAT tests for the differential association between the concepts ``male" ($A$) and ``female" ($B$) and the attributes ``career" ($X$) and ``family" ($Y$). Generally, experts in social psychology and cognitive science select stimuli that are typically representative of various concepts. In this case, $A$ contains embeddings for verbal stimuli such as ``boy," ``father," and ``man," while $X$ contains embeddings for verbal stimuli like ``office" and ``business." These linguistic, visual, and sometimes auditory stimuli are proxies for the aggregate representation of a concept in cognition. Embedding association tests use these unambiguous stimuli as semantic representations to study biased associations between the concepts being represented. Since the stimuli are chosen by experts to most accurately represent concepts, they are not polysemous or ambiguous tokens. We use these expert-selected stimuli as the basis for our tests in the image domain.

The test statistic measures the differential association of the target concepts $X$ and $Y$ with the attributes $A$ and $B$
\begin{align}s(X, Y, A, B) = \sum_{x\in X} s(x, A, B) - \sum_{y\in Y} s(y, A, B) \tag{1}\end{align}
where $s(w, A, B)$ is the differential association of $w$ with the attributes, quantified by the cosine similarity of vectors
$$s(w, A, B) = \text{mean}_{a\in A}\cos(w, a) - \text{mean}_{b\in B}\cos(w, b)$$
We test the significance of this association with a permutation test\footnote{We use an exact, non-parametric permutation test over all possible partitions. There are no normality assumptions about the distribution of the null hypothesis.} over all possible equal-size partitions $\{(X_i, Y_i)\}_i$ of $X\cup Y$ to generate a null hypothesis as if no biased associations existed. The one-sided $p$-value measures the unlikelihood of the null hypothesis
$$p = Pr[s(X_i, Y_i, A, B) > s(X, Y, A, B)]$$
and the effect size, a standardized measure of the separation between the relative association of $X$  and $Y$ with $A$ and $B$, is
$$d = \frac{\text{mean}_{x\in X}s(x, A, B) - \text{mean}_{y\in Y}s(y, A, B)}{\text{std}_{w\in X\cup Y}s(w, A, B)}$$

\noindent A larger effect size indicates a larger differential association; for instance, the large effect size $d$ in Table~\ref{tab:results} for the gender-career bias example above indicates that in human respondents, ``male" is strongly associated with ``career" attributes compared to ``female," which is strongly associated with ``family" attributes. Note that these effect sizes cannot be directly compared to effect sizes in human IATs, but the significance levels \emph{are} uniformly high. Human IATs measure individual people's associations; embedding association tests measure the aggregate association in the representation space learned from the training set. In general, significance increases with the number of stimuli; an insignificant result does not necessarily indicate a lack of bias.

One important assumption of the iEAT is that categories can be meaningfully represented by groups of images, such that the association bias measured refers to the categories of interest and not some other, similar-looking categories. Thus, a positive test result indicates only that there is an association bias between the corresponding samples' sets of target images and attribute images. To generalize to associations between abstract social concepts requires that the samples adequately represent the categories of interest. Section~\ref{sec:data} details our procedure for selecting multiple, representative stimuli, following validated approaches from prior work \cite{Greenwald1998MeasuringTest}. 

We use an adapted version of \citet{May2019OnEncoders}'s Python WEAT implementation. All code, pre-trained models, and data used to produce the figures and results in this paper can be accessed at \github.

\section{Computer Vision Models}
\label{sec:model}
To explore what kinds of biases may be embedded in image representations generated in unsupervised settings, where class labels are not available for images, we focus on two computer vision models published in summer 2020, iGPT and SimCLRv2. We extract representations of image stimuli with these two pre-trained, unsupervised image representation models. We choose these particular models because they achieve state-of-the-art performance in \emph{linear evaluation} (a measure of the accuracy of a linear image classifier trained on embeddings from each model). iGPT is the first model to learn from pixel co-occurrences to generate image samples and perform image completion tasks.

\subsubsection{Pre-training Data} 
Both models are pre-trained on ImageNet 2012, a large benchmark dataset for computer vision tasks \citep{Russakovsky2015ImageNetChallenge}.\footnote{Both models were tested on the Tensorflow version of ILSVRC 2012, available at \url{https://www.tensorflow.org/datasets/catalog/imagenet2012}.} ImageNet 2012 contains 1.2 million annotated images of 200 object classes, including a person class; even if the annotated object is not a person, a person may appear in the image. For this reason, we expect the models to be capable of generalizing to stimuli containing people  \citep{Russakovsky2013DetectingGoing, Russakovsky2015ImageNetChallenge}. While there are no publicly available pre-trained models with larger training sets, and the ``people" category of ImageNet is no longer available, this dataset is a widely used benchmark containing a comprehensive sample of images scraped from the web, primarily Flickr \citep{Russakovsky2015ImageNetChallenge}. We assume that the portrayals of people in ImageNet are reflective of the portrayal of people across the web at large, but a more contemporary study is left to future work. CIFAR-100, a smaller classification database, was also used for linear evaluation and stimuli collection \citep{Krizhevsky2009LearningImages}. 

\subsubsection{Image Representations}
Both models are \emph{unsupervised}: neither use any labels during training. Unsupervised models learn to produce embeddings based on the implicit patterns in the entire training set of image features. Both models incorporate neural networks with multiple hidden layers (each learning a different level of abstraction) and a projection layer for some downstream task. For linear classification tasks, features can be drawn directly from layers in the base neural network. As a result, there are various ways to extract image representations, each encoding a different set of information. We follow \citet{Chen2020GenerativePixels} and \citet{Chen2020ARepresentations} in choosing the features for which linear evaluation scores are highest such that the features extracted contain high-quality, general-purpose information about the objects in the image. Below, we describe the architecture and feature extraction method for each model.

\subsection{iGPT}
\label{sec:gpt}
The Image Generative Pre-trained Transformer (iGPT) model is a novel, NLP-inspired approach to unsupervised image representation. We chose iGPT for its high linear evaluation scores, minimalist architecture, and strong similarity to GPT-2 \citep{Radford2019LanguageLearners}, a transformer-based architecture that has found great success in the language domain. Transformers learn patterns in the way individual tokens in an input sequence appear with other tokens in the sequence \citep{Vaswani2017AttentionNeed}. 
\citet{Chen2020GenerativePixels} apply a structurally simple, highly parameterized version of the GPT-2 generative language pre-training architecture \citep{Radford2019LanguageLearners} to the image domain for the first time. GPT-2 uses the ``contextualized embeddings" learned by a transformer to predict the next token in a sequence and generate realistic text \citep{Radford2019LanguageLearners}. Rather than autoregressively predict the next entry in a sequence of tokens as GPT-2 does, iGPT predicts the next entry in a flattened sequence of pixels. iGPT is trained to autoregressively complete cropped images, and feature embeddings extracted from the model can be used to train a state-of-the-art linear classifier \citep{Chen2020GenerativePixels}. 

We use the largest open-source version of this model, iGPT-L 32x32, with $L=48$ layers and embedding size $1536$. All inputs are restricted to 32x32 pixels; the largest model, which takes 64x64 input, is not available to the public. Original code and checkpoints for this model were obtained from its authors at \href{https://github.com/openai/image-gpt}{github.com/openai/image-gpt}. iGPT is composed of $L$ blocks
\begin{align*}
    n^l &= \text{layer\_norm}(h^l)\\
    a^l &= h^l + \text{multihead\_attention}(n^l)\\
    h^{l+1} &= a^l+\text{mlp}(\text{layer\_norm}(a^l))
\end{align*}
where $h^l$ is the input tensor to the $l$\textsuperscript{th} block. In the final layer, called the \emph{projection head}, \citet{Chen2020GenerativePixels} learn a projection from $n^L = \text{layer\_norm}(h^L)$ to a set of logits parameterizing the conditional distributions across the sequence dimension. Because this final layer is designed for autoregressive pixel prediction, the final layer may not contain the optimal representations for object recognition tasks. \citet{Chen2020GenerativePixels} obtain the best linear classification results using embeddings extracted from a middle layer - specifically, somewhere near the 20th layer \citep{Chen2020GenerativePixels}. A linear classifier trained on these features is much more accurate than one trained on the next-pixel embeddings \citep{Chen2020GenerativePixels}. Such ``high-quality" features from the middle of the network $f^l$ are obtained by average-pooling the layer norm across the sequence dimension:
\begin{align}f^l=\langle n_i^l\rangle_i \tag{2}\end{align}
\citet{Chen2020GenerativePixels} then learn a set of \emph{class} logits from $f^l$ for their fine-tuned, supervised linear classifier, but we will just use the embeddings $f^{20}$. In general, we prefer these embeddings over embeddings from other layers for two reasons: 1) they can be more closely compared to the SimCLRv2 embeddings, which are also optimal for fine-tuning a linear classifier; 2) we hypothesize that embeddings with higher linear evaluation scores will also be more likely to embed biases, since stereotypical portrayals typically incorporate certain objects and scenes (e.g. placing men with sports equipment). In 
Appendix~\ref{appendix:logit},
we try another embedding extraction strategy and show that this hypothesis is correct.

\subsection{SimCLR}
\label{sec:simclr}
The Simple Framework for Contrastive Learning of Visual Representations (SimCLR) \citep{Chen2020ARepresentations, Chen2020BigLearners} is another state-of-the-art unsupervised image classifier. We chose SimCLRv2 because it has a state-of-the-art open source release and for variety in architecture: unlike iGPT, SimCLRv2 utilizes a traditional neural network for image encoding, ResNet \citep{He2016DeepRecognition}. SimCLRv2 extracts representations in three stages: 1) data augmentation (random cropping, random color distortions, and Gaussian blur); 2) an encoder network, ResNet \citep{He2016DeepRecognition}; 3) mapping to a latent space for contrastive learning, which maximizes agreement between the different augmented views \citep{Chen2020ARepresentations}. These representations can be used to train state-of-the-art linear image classifiers \citep{Chen2020ARepresentations, Chen2020BigLearners}. We use the largest pre-trained open-source version (the model with the highest linear evaluation scores) of SimCLRv2 \citep{Chen2020BigLearners}, obtained from its authors at \href{https://github.com/google-research/simclr}{github.com/google-research/simclr}. This pre-trained model uses a 50-layer ResNet with width $3\times$ and selective kernels (which have been shown to increase linear evaluation accuracy), and it was also pre-trained on ImageNet \citep{Russakovsky2015ImageNetChallenge}.

As with iGPT, we extract the embeddings identified by \citet{Chen2020ARepresentations} as ``high-quality" features for linear evaluation. Following \citep{Chen2020ARepresentations}, let $\tilde{x}_i$ and $\tilde{x}_j$ be two data augmentations (random cropping, random color distortion, and random Gaussian blur) of the same image. The base encoder network $f(\cdot)$ is a network of $L$ layers 
\begin{align}
h_i=f(\tilde{x}_i)=\text {ResNet}(\tilde{x}_i) \tag{3}
\end{align}
where $h_i \in \mathbb{R}^d$ is the output after the average pooling layer. During pre-training, SimCLRv2 utilizes an additional layer: a projection head $g(\cdot)$ that maps $h_i$ to a latent space for contrastive loss. The contrastive loss function can be found in \citep{Chen2020ARepresentations}.

After pre-training, \citet{Chen2020ARepresentations} discard the projection head $g(\cdot)$, using the average pool output $f(\cdot)$ for linear evaluation. Note that the projection head $g(h)$ is still necessary for pre-training high-quality representations (it improves linear evaluation accuracy by over 10\%); but \citet{Chen2020ARepresentations} find that training on $h$ rather than $z=g(h)$ also improves linear evaluation accuracy by more than 10\%.
We follow suit, using $h_i$ (the average pool output of ResNet) to represent our image stimuli, which has dimensionality $2,048$. High dimensionality is not a great obstacle;  association tests have been used with embeddings as large as $4,096$ dimensions  \citep{May2019OnEncoders}.

\section{Stimuli}
\label{sec:data}
To replicate the IATs, we systematically compiled a representative set of image stimuli for each of the concepts, or categories, listed in Table~\ref{tab:results}. Rather than attempting to specify and justify new constructs, we adhere as closely as possible to stimuli defined and employed by well-validated psychological studies. For each category (e.g. ``male" or ``science") in each IAT (e.g. Gender-Science), we drew representative images from either 1) the original IAT stimuli, if the IAT used picture stimuli \citep{Nosek2007PervasivenessStereotypes}, 2) the CIFAR-100 dataset \citep{Krizhevsky2009LearningImages}, or 3) a Google Image Search. 

This section describes how we obtained a set of images that meaningfully represent some target concept (e.g. ``male") or attribute (e.g. ``science") as it is normally, or predominantly, portrayed in society and on the web. We follow the stimuli selection criteria outlined in foundational prior work to collect the most typical and accurate exemplars \citep{Greenwald1998MeasuringTest, Greenwald2003UnderstandingAlgorithm}. For picture-IATs with readily available image stimuli, we accept those stimuli as representative and exactly replicate the IAT conditions, with two exceptions: 1) the weapon-tool IAT picture stimuli include outdated objects (e.g. cutlass, Walkman), so we chose to collect an additional, modernized set of images; 2) the disability IAT utilizes abstract symbols, so we collected a replacement set of images of real people for consistency with the training set. For IATs with verbal stimuli, we use Google Image Search as a proxy for the predominant portrayal of words (expressed as search terms) on the web (described in Section~\ref{sec:search}). Human IATs employ the same philosophy: for example, the Gender-Science IAT uses common European American names to represent male and female, because the majority of names in the U.S. are European American \citep{Nosek2002HarvestingSite}. We follow the same approach in replicating the human IATs for machines in the vision domain.

One consequence of the stimuli collection approach outlined in Section~\ref{sec:search} is that our test set will be biased towards certain demographic groups, just as the Human IATs are biased towards European American names. For example, \citet{Kay2015UnequalOccupations} showed that in 2015, search results for powerful occupations like CEO systematically under-represented women. In a case like this, we would expect to underestimate bias towards minority groups. For example, since we expect Gender-Science biases to be higher for non-White women, a test set containing more White women than non-White would exhibit lower overall bias than a test set containing an equal number of stimuli from white and non-White women. Consequently, tests on Google Image Search stimuli would be expected to result in under-estimated stereotype-congruent bias scores. While under-representation in the test set does not pose a major issue for measuring normative concepts, we cannot use the same datasets to test for intersectional bias. For those iEATs, we collected separate, equal-sized sets of images with search terms based on the categories White male, White female, Black male, and Black female, since none of the IATs specifically target these intersectional groups.

\subsection{Verbal to Image Stimuli}
\label{sec:search}
One key challenge of our approach is representing social constructs and abstract concepts such as ``male" or ``pleasantness" in images. A Google Image Search for ``pleasantness" returns mostly cartoons and pictures of the word itself. We address this difficulty by adhering as closely as possible to the verbal IAT stimuli, to ensure the validity of our replication. In verbal IATs, this is accomplished with ``buckets" of verbal exemplars that include a variety of common-place and easy-to-process realizations of the concept in question. For example, in the Gender-Science IAT, the concept ``male" is defined by the verbal stimuli ``man," ``son," ``father," ``boy," ``uncle," ``grandpa," ``husband," and ``male" \citep{Xu2014DataWebsite}. To closely match the representations tested by these IATs, we use these sets of words to search for substitute image stimuli that portray one of these words or phrases.  For the vast majority of exemplars, we were able to find direct visualizations of the stimuli as an isolated person, object, or scene. For example, Figure~\ref{fig:diagram} depicts sample image stimuli corresponding to the verbal stimuli ``orchid" (for category ``flower"), ``centipede" (``insect"), ``sunset" (``pleasant"), and ``morgue" (``unpleasant").\footnote{In the original IATs, the category set sizes $N_t$ and $N_a$ range from 5-15 exemplars. We collected $n\approx5$ images for each exemplar such that $N_t$ and $N_a$ are 30-50.  Significance could be increased by including more stimuli, at the risk of diluting the test set with less-representative images from farther down in the search results.}

We collected images for each verbal stimulus from either CIFAR-100\footnote{We first check for test images in CIFAR-100 because iGPT performs well in out-of-sample linear evaluation on this dataset \citep{Chen2020ARepresentations}.} or Google Image Search according to a systematic procedure detailed in Appendix~\ref{appendix:stimuli}. This procedure controls for image characteristics that might confound the category we are attempting to define (e.g. lighting, background, dominant colors, placement) in several ways: 1) we collected more than one for each verbal stimulus, in case of idiosyncrasies in the images collected; 2) for stimuli referring to an object or person, we chose images that isolated the object or person of interest against a plain background, unless the object filled the whole image;
3) when an attribute stimulus refers to a group of people, we chose only images where the target concepts were evenly represented in the attribute images;\footnote{For example, for the ``family" attribute in the Gender-Career test, we chose only images of families with equal numbers of men and women.} 4) for the picture-IATs, we accepted the original image stimuli to exactly reconstruct the original test conditions. We also did not alter the original verbal stimuli, relying instead on the construct validity of the original IAT experiments.\footnote{One exception: the Gender-Career IAT used specific male- and female-sounding names, rather than general exemplars like ``man" or ``father" as in the Gender-Science IAT. We use the general exemplars for both tests.} For each verbal stimulus, Appendix~\ref{appendix:stimuli} lists corresponding search terms and the precise number of images collected. All the images used to represent the concepts being tested are available at \github.

\subsection{Choosing Valence Stimuli}
Valence, the intrinsic pleasantness or goodness of things, is one of the principal dimensions of affect and cognitive heuristics that shape attitudes and biases \citep{Greenwald1998MeasuringTest}. Many IATs quantify implicit bias by comparing two social groups to the valence attributes ``pleasant" vs. ``unpleasant." Here, positive valence will denote ``pleasantness" and negative valence will denote ``unpleasantness." The verbal exemplars for valence vary slightly from test to test. Rather than create a new set of image stimuli for each valence IAT, we collected one, large consolidated set from an experimentally validated database \citep{Bellezza1986WordsStudents} of low and high valence words (e.g. ``rainbow," ``morgue") commonly used in the valence IATs. To quantify norms, \citep{Bellezza1986WordsStudents} asked human participants to rate these non-social words for ``pleasantness" and ``imagery" in a controlled laboratory setting. Because some of the words for valence do not correspond to physical objects, we collected images for verbal stimuli with high valence and imagery scores. We used the same procedure as for all the other verbal stimuli (described above in Section~\ref{sec:search}). The full list of verbal valence stimuli can be found in Appendix~\ref{appendix:words}.

\section{Evaluation}
\label{sec:evaluation}

We evaluate the validity of iEAT by comparing the results to human and natural language biases measured in prior work. We obtain stereotype-congruent results for baseline, or ``universal," biases. We also introduce a simple experiment to test how often the iEAT incorrectly finds bias in a random set of stimuli.

\textbf{Predictive Validity.} We posit that iEAT results have predictive validity if they correspond to ground-truth IAT results for humans or WEAT results in word embeddings. In this paper, we validate the iEAT by replicating several human IATs as closely as possible (as described in Section~\ref{sec:data}) and comparing the results. We find that embeddings extracted from at least one of the two models we test display significant bias for 8 of the 15 ground-truth human IATs we replicate (Section~\ref{sec:results}). The insignificant biases are likely due to small sample sizes. We also find evidence supporting each of the intersectional hypotheses listed in Section~\ref{subsec:intersectional}, which have also been empirically validated in a study with human participants \citep{Ghavami2013AnHypotheses}.

\textbf{Baselines.} As a baseline, we replicate a ``universal" bias test presented in the first paper introducing the IAT \citep{Greenwald1998MeasuringTest}: the association between flower vs. insects and pleasant vs. unpleasant. If human-like biases are encoded in unsupervised image models, we would expect a strong and statistically significant flower-insect valence bias, for two reasons: 1) as \citet{Greenwald1998MeasuringTest} conjecture, this test measures a close-to-universal baseline human bias; 2) our models (described in Section~\ref{sec:model}) achieve state-of-the-art performance when classifying simple objects including flowers and bees.\footnote{A linear image classifier trained on iGPT embeddings reaches 88.5\% accuracy on CIFAR-100; SimCLRv2 embeddings reach 89\% accuracy \citep{Chen2020GenerativePixels}.} The presence of universal bias and absence of random bias suggests our conclusions are valid for other social biases.

\textbf{Specificity.} Prior work on embedding association tests does not evaluate the false positive rate. To validate the specificity of our significance estimation, we created 1,000 random partitions of $X\cup Y\cup A\cup B$ from the flower-insect test to evaluate true positive detection. Our false positive rate is roughly bounded by the $p$-value: 10.3\% of these random tests resulted in a false positive at $p<10^{-1}$; 1.2\% were statistically significant false positives at $p<10^{-2}$.

\section{Experiments and Results}
\label{sec:results}
In correspondence with the human IAT, we find several significant racial biases and gender stereotypes, including intersectional biases, shared by both iGPT and SimCLRv2 when pre-trained on ImageNet.

\subsection{iEATs}
Effect sizes and $p$-values from the permutation test for each bias type measurement are reported in Table~\ref{tab:results} and interpreted below.
\subsubsection{Widely Accepted Biases}
First, we apply the iEAT to the widely accepted baseline Insect-Flower IAT, which measures the association of insects and flowers with pleasantness and unpleasantness, respectively. As hypothesized, we find that embeddings from both models contain significant positive biases in the same direction as the human participants, associating flowers with pleasantness and insects with unpleasantness, with $p<10{-1}$ (Table~\ref{tab:results}). Notably, the magnitude of bias is greater for SimCLRv2 (effect size $1.69$, $p<10^{-3}$) than for iGPT (effect size $0.34$, $p<10^{-1}$). In general, SimCLRv2 embeddings contain stronger biases than iGPT embeddings but do not contain as many kinds of bias. We conjecture that because SimCLRv2 transforms images before training (including color distortion and blurring) and is more architecturally complex than iGPT \citep{Chen2020ARepresentations}, its embeddings become more suitable for concrete object classification as opposed to implicit social patterns.

\subsubsection{Racial Biases}
Both models display statistically significant racial biases, including both valence and stereotype biases. The racial attitude test, which measures the differential association of images of European Americans vs. African Americans with pleasantness and unpleasantness, shows no significant biases. But embeddings extracted from both models exhibit significant bias for the Arab-Muslim valence test, which measures the association of images of Arab-Americans vs. others with pleasant vs. unpleasant images. Also, embeddings extracted with iGPT exhibit strong bias large effect size (effect size $1.26$, $p<10{-2}$) for the Skin Tone test, which compares valence associations with faces of lighter and darker skin tones. These findings relate to anecdotal examples of software that claim to make faces more attractive by lightening their skin color. Both iGPT and SimCLRv2 embeddings also associate White people with tools and Black people with weapons in both classical and modernized versions of the Weapon IAT.

\begin{table*}[ht!]
    \centering
    \caption{iEAT tests for the association between target concepts $X$ vs. $Y$ (represented by $n_t$ images each) and attributes $A$ vs. $B$ (represented by $n_a$ images each) in embeddings generated by an unsupervised model. Effect sizes $d$ represent the magnitude of bias, colored by conventional small (0.2), medium (0.5), and large (0.8). Permutation $p$-values indicate significance. Reproduced from \citet{Nosek2007PervasivenessStereotypes}, the original human IAT effect sizes are all statistically significant with $p<10^{-8}$; they can be compared to our effect sizes in sign but not in magnitude.}
    \label{tab:results}
    \begin{small}
        \begin{tabular}{llllllllllr}
\toprule
{} &                $X$ &               $Y$ &       $A$ &           $B$ & $n_t$ & $n_a$ &   Model &         iEAT              $d$ &     iEAT    $p$ &                  IAT $d$ \\
\midrule
Age\textsuperscript{\textdagger}        &              Young &               Old &  Pleasant &    Unpleasant &     6 &    55 &    iGPT &   \cellcolor{d_small}0.42 &        0.24 &  \cellcolor{d_large}1.23 \\
                                        &                    &                   &           &               &       &       &  SimCLR &  \cellcolor{d_medium}0.59 &        0.16 &  \cellcolor{d_large}1.23 \\
Arab-Muslim                             &              Other &       Arab-Muslim &  Pleasant &    Unpleasant &    10 &    55 &    iGPT &   \cellcolor{d_large}0.86 &        0.02 &  \cellcolor{d_small}0.33 \\
                                        &                    &                   &           &               &       &       &  SimCLR &   \cellcolor{d_large}1.06 &  $<10^{-2}$ &  \cellcolor{d_small}0.33 \\
Asian\textsuperscript{$\mathsection$}   &  European American &    Asian American &  American &       Foreign &     6 &     6 &    iGPT &   \cellcolor{d_small}0.25 &        0.34 & \cellcolor{d_medium}0.62 \\
                                        &                    &                   &           &               &       &       &  SimCLR &   \cellcolor{d_small}0.47 &        0.21 & \cellcolor{d_medium}0.62 \\
Disability\textsuperscript{\textdagger} &           Disabled &             Abled &  Pleasant &    Unpleasant &     4 &    55 &    iGPT &                     -0.02 &        0.53 &  \cellcolor{d_large}1.05 \\
                                        &                    &                   &           &               &       &       &  SimCLR &   \cellcolor{d_small}0.38 &        0.34 &  \cellcolor{d_large}1.05 \\
Gender-Career                           &               Male &            Female &    Career &        Family &    40 &    21 &    iGPT &  \cellcolor{d_medium}0.62 &  $<10^{-2}$ &   \cellcolor{d_large}1.1 \\
                                        &                    &                   &           &               &       &       &  SimCLR &  \cellcolor{d_medium}0.74 &  $<10^{-3}$ &   \cellcolor{d_large}1.1 \\
Gender-Science                          &               Male &            Female &   Science &  Liberal Arts &    40 &    21 &    iGPT &   \cellcolor{d_small}0.44 &        0.02 &  \cellcolor{d_large}0.93 \\
                                        &                    &                   &           &               &       &       &  SimCLR &                     -0.10 &        0.67 &  \cellcolor{d_large}0.93 \\
Insect-Flower                           &             Flower &            Insect &  Pleasant &    Unpleasant &    35 &    55 &    iGPT &   \cellcolor{d_small}0.34 &        0.07 &  \cellcolor{d_large}1.35 \\
                                        &                    &                   &           &               &       &       &  SimCLR &   \cellcolor{d_large}1.69 &  $<10^{-3}$ &  \cellcolor{d_large}1.35 \\
Native\textsuperscript{$\mathsection$}  &  European American &   Native American &      U.S. &         World &     8 &     5 &    iGPT &                     -0.33 &        0.73 &  \cellcolor{d_small}0.46 \\
                                        &                    &                   &           &               &       &       &  SimCLR &                     -0.19 &        0.65 &  \cellcolor{d_small}0.46 \\
Race\textsuperscript{\textdagger}       &  European American &  African American &  Pleasant &    Unpleasant &     6 &    55 &    iGPT &                     -0.62 &        0.85 &  \cellcolor{d_large}0.86 \\
                                        &                    &                   &           &               &       &       &  SimCLR &                     -0.57 &        0.83 &  \cellcolor{d_large}0.86 \\
Religion                                &       Christianity &           Judaism &  Pleasant &    Unpleasant &     7 &    55 &    iGPT &   \cellcolor{d_small}0.37 &        0.25 &                    -0.34 \\
                                        &                    &                   &           &               &       &       &  SimCLR &   \cellcolor{d_small}0.36 &        0.26 &                    -0.34 \\
Sexuality                               &                Gay &          Straight &  Pleasant &    Unpleasant &     9 &    55 &    iGPT &                     -0.03 &        0.52 & \cellcolor{d_medium}0.74 \\
                                        &                    &                   &           &               &       &       &  SimCLR &                      0.04 &        0.47 & \cellcolor{d_medium}0.74 \\
Skin-Tone\textsuperscript{\textdagger}  &              Light &              Dark &  Pleasant &    Unpleasant &     7 &    55 &    iGPT &   \cellcolor{d_large}1.26 &  $<10^{-2}$ & \cellcolor{d_medium}0.73 \\
                                        &                    &                   &           &               &       &       &  SimCLR &                     -0.19 &        0.71 & \cellcolor{d_medium}0.73 \\
Weapon\textsuperscript{$\mathsection$}  &              White &             Black &      Tool &        Weapon &     6 &     7 &    iGPT &   \cellcolor{d_large}0.86 &        0.07 &   \cellcolor{d_large}1.0 \\
                                        &                    &                   &           &               &       &       &  SimCLR &   \cellcolor{d_large}1.38 &  $<10^{-2}$ &   \cellcolor{d_large}1.0 \\
Weapon (Modern)                         &              White &             Black &      Tool &        Weapon &     6 &     9 &    iGPT &   \cellcolor{d_large}0.88 &        0.06 &                      N/A \\
                                        &                    &                   &           &               &       &       &  SimCLR &   \cellcolor{d_large}1.28 &        0.01 &                      N/A \\
Weight\textsuperscript{\textdagger}     &               Thin &               Fat &  Pleasant &    Unpleasant &    10 &    55 &    iGPT &   \cellcolor{d_large}1.67 &  $<10^{-3}$ &  \cellcolor{d_large}1.83 \\
\multicolumn{7}{l}{\textsuperscript{$\mathsection$}\footnotesize{ Originally a picture-IAT (image-only stimuli). \textsuperscript{\textdagger} Originally a mixed-mode IAT (image and verbal stimuli).}} &  SimCLR &                     -0.30 &        0.74 &  \cellcolor{d_large}1.83 \\
\bottomrule
\end{tabular}
    \end{small}
\end{table*}

\subsubsection{Gender Biases}
There are statistically significant gender biases in both models, though not for both stereotypes we tested. In the Gender-Career test, which measures the relative association of the category ``male" with career attributes like ``business" and ``office" and the category ``female" with family-related attributes like ``children" and ``home," embeddings extracted from both models exhibit significant bias (iGPT effect size $0.62$, $p<10^{-2}$, SimCLRv2 effect size $0.74$, $p<10^{-3}$). This finding parallels \citet{Kay2015UnequalOccupations}'s observation that image search results for powerful occupations like CEO systematically under-represented women. In the Gender-Science test, which measures the association of ``male" with ``science" attributes like math and engineering and ``female" with ``liberal arts" attributes like art and writing, only iGPT displays significant bias (effect size $0.44$, $p<10^{-1}$). 

\subsubsection{Other Biases}
For the first time, we attempt to replicate several other tests measuring weight stereotypes and attitudes towards the elderly or people with disabilities. iGPT displays an additional bias (effect size $1.67$, $p=10^{-4}$) towards the association of thin people with pleasantness and overweight people with unpleasantness. We found no significant bias for the Native American or Asian American stereotype tests, the Disability valence test, or the Age valence test. For reference, significant age biases have been detected in static word embeddings; the others have not been tested because they use solely image stimuli \citep{Caliskan2017}. Likely, the target sample sizes for these tests are too low; all three of these tests use picture stimuli from the original IAT, which are all limited to fewer than 10 images. Replication with an augmented test set is left to future work. Note that lack of significance in a test, even if the sample size is sufficiently large, does not indicate the embeddings from either model are definitively bias-free. While these tests did not \textit{confirm} known human biases regarding foreigners, people with disabilities, and the elderly, they also did not \textit{contradict} any known human-like biases.

\subsection{Intersectional Biases}
\begin{table*}[ht!]
    \centering
    \caption{iEAT tests for the association between intersectional group $X$ vs. $Y$ (represented by $n_t$ images each) and attributes $A$ vs. $B$ (represented by $n_a$ images each) in embeddings produced by an unsupervised model. Effect sizes $d$ represent the magnitude of bias, colored by conventional small (0.2), medium (0.5), and large (0.8). Permutation $p$-values indicate significance.}
    \label{tab:intersectional}
    \begin{small}
    \begin{tabular}{lllllllll}
\toprule
{} &           $X$ &           $Y$ &       $A$ &           $B$ & $n_t$ & $n_a$ &                       $d$ &         $p$ \\
\midrule
Gender-Career (MF)    &          Male &        Female &    Career &        Family &    40 &    21 &   \cellcolor{d_large}0.81 &  $<10^{-3}$ \\
Gender-Career (WMBF)  &    White Male &  Black Female &           &               &    20 &    21 &                      0.20 &        0.27 \\
Gender-Career (WMBM)  &    Black Male &    White Male &    Career &        Family &    20 &    21 &   \cellcolor{d_large}0.89 &  $<10^{-2}$ \\
Gender-Career (WMWF)  &    White Male &  White Female &           &               &    20 &    21 &   \cellcolor{d_large}0.97 &  $<10^{-3}$ \\
Gender-Science (MF)   &          Male &        Female &   Science &  Liberal Arts &    40 &    21 &                     0.00 &        0.50 \\
Gender-Science (WMBF) &    White Male &  Black Female &           &               &    20 &    21 &  \cellcolor{d_medium}0.80 &  $<10^{-2}$ \\
Gender-Science (WMBM) &    White Male &    Black Male &   Science &  Liberal Arts &    20 &    21 &   \cellcolor{d_small}0.49 &        0.06 \\
Gender-Science (WMWF) &    White Male &  White Female &           &               &    20 &    21 &                     -0.37 &        0.88 \\
Valence (BFBM)        &  Black Female &    Black Male &  Pleasant &    Unpleasant &    20 &    55 &                      0.17 &        0.29 \\
Valence (BW)          &         White &         Black &           &               &    40 &    55 &   \cellcolor{d_large}1.16 &  $<10^{-3}$ \\
Valence (FM)          &        Female &          Male &  Pleasant &    Unpleasant &    40 &    55 &   \cellcolor{d_small}0.39 &        0.04 \\
Valence (WFBF)        &  White Female &  Black Female &           &               &    20 &    55 &   \cellcolor{d_large}1.51 &  $<10^{-3}$ \\
Valence (WFBM)        &  White Female &    Black Male &  Pleasant &    Unpleasant &    20 &    55 &   \cellcolor{d_large}1.46 &  $<10^{-3}$ \\
Valence (WMBF)        &    White Male &  Black Female &           &               &    20 &    55 &   \cellcolor{d_large}0.83 &  $<10^{-2}$ \\
Valence (WMBM)        &    White Male &    Black Male &  Pleasant &    Unpleasant &    20 &    55 &   \cellcolor{d_large}0.88 &  $<10^{-2}$ \\
Valence (WMWF)        &  White Female &    White Male &           &               &    20 &    55 &  \cellcolor{d_medium}0.79 &  $<10^{-2}$ \\
\bottomrule
\end{tabular}
    \end{small}
\end{table*}

\subsubsection{Intersectional Valence}
Intersectional valence tests with the iGPT embeddings are the most consistent with social psychology, exhibiting results predicted by the intersectionality, race, and gender hypotheses listed in Section~\ref{sec:approach} \citep{Ghavami2013AnHypotheses}. Overall, iGPT embeddings contain a positive valence bias towards White people and a negative valence bias towards Black people (effect size $1.16$, $p<10^{-3}$), as in the human Race IAT \citep{Nosek2007PervasivenessStereotypes}. As predicted by the race hypothesis, the same bias is significant but less severe for both White males vs. Black males (iGPT effect size $0.88$, $p<10^{-2}$) and White males vs. Black females (iGPT effect size $0.83$, $p<10^{-2}$), and the White female vs. Black female bias is insignificant; in general, race biases are more similar to the race biases between men. We hypothesize that as in text corpora, computer vision datasets are dominated by the majority social groups (men and White).

As predicted by the gender hypothesis, our results also conform with the theory that females are associated with positive valence when compared to males \citep{Eagly1991AreEmotions}, but only when those groups are White (iGPT effect size $0.79$, $p<10^{-2}$); there is no significant valence bias for Black females vs. Black males. This insignificant result might be due to the under-representation of Black people in the visual embedding space. The largest differential valence bias of all our tests emerges between White females and Black males; White females are associated with pleasant valence and Black males with negative valence (iGPT effect size $1.46$, $p<10^{-3}$).

\vspace{-2mm}

\subsubsection{Intersectional Stereotypes} We find significant but contradictory intersectional differences in gender stereotypes (Table~\ref{tab:intersectional}). For Gender-Career stereotypes, the iGPT-encoded bias for White males vs. Black females is insignificant though there is a bias (effect size $0.81$, $p<10^{-3}$) for male vs. female in general. There is significant Gender-Career stereotype bias between embeddings of White males vs. White females (iGPT effect size $0.97$, $p<10^{-3}$), even higher than the general case; this result conforms to the race hypothesis, which predicts gender stereotypes are more similar to the stereotypes between Whites than between Blacks. The career-family bias between White males and Black males is reversed; embeddings for images of Black males are more associated with career and images of White men with family (iGPT effect size $0.89$, $p<10^{-2}$). One explanation for this result is under-representation; there are likely fewer photos depicting Black men with non-stereotypical male attributes.

Unexpectedly, the intersectional test of male vs. female (with equal representation for White and Black people) reports no significant Gender-Science bias, though the normative test (with unequal representation) does (Table~\ref{tab:results}). Nevertheless, race-science stereotypes do emerge when White males are compared to Black males (iGPT effect size $0.49$, $p<10^{-1}$) and, to an even greater extent, when White males are compared to Black females (iGPT effect size $0.80$, $p<10^{-2}$), confirming the intersectional hypothesis \citep{Ghavami2013AnHypotheses}. But visual Gender-Science biases do not conform to the race hypothesis; the gender stereotype between White males and White females is insignificant, though the overall male vs. female bias is not.

\subsection{Origins of Bias}
\subsubsection{Bias in Web Images}
Do these results correspond with our hypothesis that biases are learned from the co-occurrence of social group members with certain stereotypical or high-valence contexts? Both our models were pre-trained on ImageNet, which is composed of images collected from Flickr and other Internet sites \citep{Russakovsky2015ImageNetChallenge}. \citet{Yang2020TowardsHierarchy} show that the ImageNet categories unequally represent race and gender; for instance, the ``groom" category may contain mostly White people. Under-representation in the training set could explain why, for instance, White people are more associated with pleasantness and Black people with unpleasantness. There is a similar theory in social psychology: most bias takes the form of in-group favoritism, rather than out-group derogation \citep{Hewstone2002IntergroupBias}. In image datasets, favoritism could take the form of unequal representation and have similar effects. For example, one of the exemplars for ``pleasantness" is ``wedding," a positive-valence, high imagery word \citep{Bellezza1986WordsStudents}; if White people appear with wedding paraphernalia more often than Black people, they could be automatically associated with a concept like ``pleasantness," even though no explicit labels for ``groom" and ``White" are available during training.

Likewise, the portrayal of different social groups in context may be automatically learned by unsupervised image models. \citet{Wang2020REVISE:Datasets} find that in OpenImages (also scraped from Flickr) \citep{Kuznetsova2018TheScale}, a similar benchmark classification dataset, a higher proportion of ``female" images are set in the scene ``home or hotel" than ``male" images. ``male" is more often depicted in ``industrial and construction" scenes. This difference in portrayal could account for the Gender-Career biases embedded in unsupervised image embeddings. In general, if the portrayal of people in Internet images reflects human social biases that are documented in cognition and language, we conclude that unsupervised image models could automatically learn human-like biases from large collections of online images.

\subsubsection{Bias in Autoregression}
\begin{figure}[!t]
    \centering
    \begin{subfigure}[t]{\linewidth}
        \centering
        
        \includegraphics[width=0.15\textwidth]{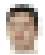}
         
        \caption{Cropped image of an artificial \citep{2021GeneratedPhotos}, white- male-passing face.}
    \end{subfigure}
    \quad
    \begin{subfigure}[t]{\linewidth}
        \centering
        \includegraphics[width=\textwidth]{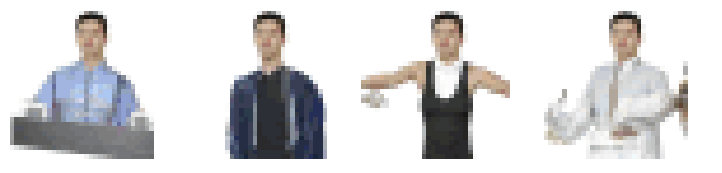}
        \includegraphics[width=\textwidth]{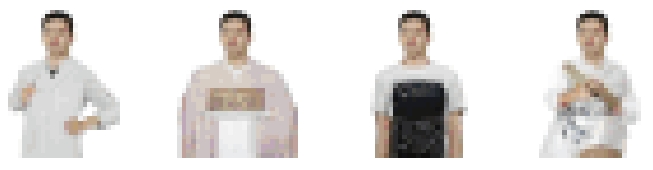}
        \caption{8 random autoregressive completions of the cropped images. 6 depict career-related attire.}
    \end{subfigure}
    \caption{Example of career associations in image completion of a male face with iGPT, pre-trained on ImageNet.}
    \label{fig:completion}
\end{figure}

Though the next-pixel prediction features contained very little significant bias, they may still propagate stereotypes in practice. For example, the incautious and unethical application of a generative model like iGPT could produce biased depictions of people. As a qualitative case study, we selected 5 male- and 5 female-appearing artificial faces from a database \citep{2021GeneratedPhotos} generated with StyleGAN \citep{Karras2019ANetworks}. We decided to use images of non-existent people to avoid perpetuating any harm to real individuals. We cropped the portraits below the neck and used iGPT to generate 8 different completions (with the temperature hyperparameter set to $1.0$, following \citet{Chen2020GenerativePixels}). We found that completions of woman \emph{and} men are often sexualized: for female faces, 52.5\% of completions featured a bikini or low-cut top; for male faces, 7.5\% of completions were shirtless or wore low-cut tops, while 42.5\% wore suits or other career-specific attire. One held a gun. This behavior might result from the sexualized portrayal of people, especially women, in internet images \citep{Graff2013Low-cutGirls} and serves as a reminder of computer vision's controversial history with Playboy centerfolds and objectifying images \citep{Iozzio2016TheResearch}. To avoid promoting negative biases, Figure~\ref{fig:completion} shows only an example of male-career associations in completions of a GAN-generated face.

\section{Discussion} 
\label{sec:discussion}
By testing for bias in unsupervised models pre-trained on a widely used large computer vision dataset, we show how biases may be learned automatically from images and embedded in general-purpose representations. Not only do we observe human-like biases in the majority of our tests, but we also detect 4 of the 5 human biases replicated in natural language \citep{Caliskan2017}. \citet{Caliskan2017} show that artifacts of the societal status quo, such as occupational gender statistics, are imprinted in online text and mimicked by machines. We suggest that a similar phenomenon is occurring for online images. One possible culprit is confirmation bias \citep{Schweiger2014ConfirmationEvaluation}, the tendency of individuals to consume and produce content conforming to group norms. Self-supervised models exhibit the same tendency \citep{Arazo2020Pseudo-LabelingLearning}.

In addition to confirming human and natural language machine biases in the image domain, the iEAT measures visual biases that may implicitly affect humans and machines but cannot be captured in text corpora. \citet{Foroni2010Picture-IATIAT} conjecture that in humans, picture-IATs and word-IATs measure different mental processes. More research is needed to explore biases embedded in images and investigate their origins, as \citet{Brunet2019UnderstandingEmbeddings} suggest for language models. \citet{Tenney2019WhatRepresentations} show that contextual representations learn syntactic and semantic features from the context. \citet{Voita2019TheObjectives} explain the change of vector representations among layers based on the compression/prediction trade-off perspective. Advances in this direction would contribute to our understanding of the causal factors behind visual perception and biases related to cognition and language acquisition.

Our methods come with some limitations. The biases we measure are in large part due to patterns learned from the pre-training data, but ImageNet 2012 does not necessarily represent the entire population of images currently produced and circulated on the Internet. Additionally, ImageNet 2012 is intended for object detection, not distinguishing people's social attributes, and both our models were validated for non-person object classification.\footnote{Recently, \citet{Yang2020TowardsHierarchy} proposed updates to improve fairness and representation in the ImageNet ``person" category that could change our results.} The largest version of iGPT (not publicly available) was pre-trained on 100 million additional web images \citep{Chen2020GenerativePixels}. Given the financial and carbon costs of the computation required to train highly parameterized models like iGPT, we did not train our own models on larger-scale corpora. Complementary iEAT bias testing with unsupervised models pre-trained on an updated version of ImageNet could help quantify the effectiveness of dataset de-biasing strategies.

A model like iGPT, pre-trained on a more comprehensive private dataset from a platform like Instagram or Facebook, could encode much more information about contemporary social biases. Clearview AI reportedly scraped over 3 billion images from Facebook, YouTube, and millions of other sites for their face recognition model \citep{Hill2020TheIt}. \citet{Dosovitskiy2021AnScale} recently trained a very similar transformer model on Google's JFT-300M, a 300 million image dataset scraped from the web \citep{Sun2017RevisitingEra}.
Further research is needed to determine how architecture choices affect embedded biases and how dataset filtering and balancing techniques might help \citep{Wang2020TowardsMitigation, Wang2019BalancedRepresentations}. Previous metric-based and adversarial approaches generally require labeled datasets \citep{Wang2019BalancedRepresentations, Wang2020REVISE:Datasets, Wang2020TowardsMitigation}. Our method avoids the limitations of laborious manual labeling.

Though models like these may be useful for quantifying contemporary social biases as they are portrayed in vast quantities of images on the Internet, our results suggest the use of unsupervised pre-training on images at scale is likely to propagate harmful biases. Given the high computational and carbon cost of model training at scale, transfer learning with pre-trained models is an attractive option for practitioners. But our results indicate that patterns of stereotypical portrayal of social groups do affect unsupervised models, so careful research and analysis are needed before these models make consequential decisions about individuals and society. Our method can be used to assess task-agnostic biases contained in a dataset to enhance transparency \citep{Gebru2018DatasheetsDatasets, Mitchell2019ModelReporting}, but bias mitigation for unsupervised transfer learning is a challenging open problem.

\section{Conclusions}
\label{sec:conclusion}
We develop a principled method for measuring bias in unsupervised image models, adapting embedding association tests used in the language domain. With image embeddings extracted by state-of-the-art unsupervised image models pre-trained on ImageNet, we successfully replicate validated bias tests in the image domain and document several social biases, including severe intersectional bias. Our results suggest that unsupervised image models learn human biases from the way people are portrayed in images on the web. These findings serve as a caution for computer vision practitioners using transfer learning: pre-trained models may embed all types of harmful human biases from the way people are portrayed in training data, and model design choices determine whether and how those biases are propagated into harms downstream.

\begin{acks}
This material is based on research partially supported by the U.S. National Institute of Standards and Technology (NIST) Grant \\60NANB20D212. Any opinions, findings, and conclusions or recommendations expressed in this material are those of the authors and do not necessarily reflect those of NIST.
\end{acks}

\clearpage

\bibliographystyle{ACM-Reference-Format}
\bibliography{references}

\clearpage
\appendix
\section{Attribute Words}
\label{appendix:words}
We selected the following words for high/low valence and high imagery from the scores collected by \citet{Bellezza1986WordsStudents} in a laboratory experiment. A specific algorithm for systematically selecting words with high imagery and extreme valence is included in our code at \github.\\

\noindent\emph{Positive words}: baby, ocean, beach, butterfly, gold, rainbow, sunset, money, diamond, flower, sunrise\\
\noindent\emph{Negative words}: devil, morgue, slum, corpse, coffin, jail, roach, funeral, prison, vomit, crash

\section{Stimuli collection procedure}
\label{appendix:stimuli}

We collected $n$ images for each verbal stimulus using the following procedure:
\begin{enumerate}[leftmargin=*]
\item If there is a CIFAR-100 category corresponding to the stimulus, we selected a random sample of $n$ images from that category in CIFAR \citep{Krizhevsky2009LearningImages}.\footnote{Because the verbal stimuli are very specific, only 3 of over 105 IAT verbal stimuli appear in CIFAR-100; the rest were collected with Google Image Search.}
\item Otherwise, we searched for the verbal stimuli verbatim on Google Image Search in private Chrome window with SafeSearch off on September 5th, September 18th and October 1st, 2020. We accepted the first $n$ results of the search meeting the following criteria:\footnote{A few words were too abstract to be easily visualized. These words are listed in Appendix~\ref{appendix:stimuli} with a sample size of 0.}
    \begin{itemize}
        \item Includes only the object, person, or scene specified by the stimulus.\footnote{Some verbal stimuli (e.g. ``salary") are difficult to express verbally without the use of symbols (e.g. a picture of cash). In these cases, we collected only the first image ($n=1$) that meets the criteria, preferring image stimuli corresponding to other, more visual cues and representations.} 
        \item For objects and people, has a plain background, to avoid including confounding scenes or objects.\footnote{If no images with white or gray backgrounds appeared in the first 50 results, we searched for ``[stimulus] + \{white, plain\} background."}
        \item Has no watermark or other text. Watermarks and text could confound the verbal stimulus being represented.
        \item Shows a real object, person, or scene - is not a cartoon or sketch. ImageNet does not include a great quantity of cartoons or sketches, so we do not expect our models to generalize well to these kinds of objects/scenes \citep{Recht2019DoImagenet}.
    \end{itemize}
\item If no images in the first 50 results from the verbatim search met these criteria, we added a clarifying search term (e.g. ``biology lab" instead of ``biology").
\item Crop each image squarely (iGPT accepts only square images as input), centering the object or person of interest to ensure the entire object, person, or scene is included in the image.
\end{enumerate}

For every verbal stimulus used to collect image stimuli for the verbal and mixed-mode IATs, we recorded the verbal stimulus (word or phrase), search terms used to collect images, and the number of images collected in a CSV file along with our code at \github.

\section{Disparate Bias Across Model Layers}
\label{appendix:logit}
Model design choices might also have an effect on how social bias is learned in visual embeddings. We find that embedded social biases vary not only between models pre-trained on the same data but also within layers of the same model. In addition to the high quality embeddings extracted from the middle of the model, we tested embeddings extracted at the next-pixel logistic prediction layer of iGPT. This logit layer, when taken as a set of probabilities with softmax or a similar function, is used to solve the next-pixel prediction task for unconditional image generation and conditional image completion \citep{Chen2020GenerativePixels}. 

Table~\ref{tab:logit} reports the iEAT tests results for these embeddings, which did not display the same correspondence with human bias as the embeddings for image classification. We found that unlike the high quality embeddings, next-pixel prediction embeddings do not exhibit the baseline Insect-Flower valence bias and only encode significant bias at the $10^{-1}$ level for the Gender-Science and Sexuality IATs. 

To explain this difference in behavior, recall that the neural network used in iGPT learns different levels of abstraction at each layer; as an example, imagine that first layer encodes lighting particularly well, while the second layer begins to encode curves. The contradiction between biases in the middle layers and biases in the projection head are consistent with two previous findings: 1) bias is encoded disparately across the layers of unsupervised pre-trained models, as \citet{Bommasani2020InterpretingEmbeddings} show in the language domain; 2) in transformer models, the highest quality features for image classification, and possibly also social bias prediction, are found in the middle of the base network \citep{Chen2020GenerativePixels}. Evidently, bias depends not only on the training data but also on the choice of model.

\begin{table*}[ht!]
    \caption{iEAT tests for the association between target concepts $X$ vs. $Y$ (represented by $n_t$ images each) and attributes $A$ vs. $B$ (represented by $n_a$ images each) in embeddings for iGPT next-pixel prediction. Association effect sizes $d$, colored by conventional small (0.2), medium (0.5), and large (0.8) size are reported alongside permutation $p$-values.}
    \label{tab:logit}
    \begin{tabular}{lllllllll}
\toprule
{} &                $X$ &               $Y$ &       $A$ &           $B$ & $n_t$ & $n_a$ &                       $d$ &   $p$ \\
\midrule
Age\textsuperscript{\textdagger}        &              Young &               Old &  Pleasant &    Unpleasant &     6 &    55 &   \cellcolor{d_small}0.38 &  0.38 \\
Arab-Muslim                             &              Other &       Arab-Muslim &  Pleasant &    Unpleasant &    10 &    55 &                      0.06 &  0.42 \\
Asian\textsuperscript{$\mathsection$}   &  European American &    Asian American &  American &       Foreign &     6 &     6 &   \cellcolor{d_small}0.25 &  0.36 \\
Disability\textsuperscript{\textdagger} &           Disabled &             Abled &  Pleasant &    Unpleasant &     4 &    55 &                     -0.65 &  0.76 \\
Gender-Career                           &               Male &            Female &    Career &        Family &    40 &    21 &                      0.04 &  0.44 \\
Gender-Science                          &               Male &            Female &   Science &  Liberal Arts &    40 &    21 &   \cellcolor{d_small}0.37 &  0.06 \\
Insect-Flower                           &             Flower &            Insect &  Pleasant &    Unpleasant &    35 &    55 &                     -0.32 &  0.91 \\
Native\textsuperscript{$\mathsection$}  &  European American &   Native American &      U.S. &         World &     8 &     5 &   \cellcolor{d_small}0.32 &  0.26 \\
Race\textsuperscript{\textdagger}       &  European American &  African American &  Pleasant &    Unpleasant &     6 &    55 &                     -0.17 &  0.62 \\
Religion                                &       Christianity &           Judaism &  Pleasant &    Unpleasant &     7 &    55 &   \cellcolor{d_small}0.29 &  0.30 \\
Sexuality                               &                Gay &          Straight &  Pleasant &    Unpleasant &     9 &    55 &  \cellcolor{d_medium}0.69 &  0.08 \\
Skin-Tone\textsuperscript{\textdagger}  &              Light &              Dark &  Pleasant &    Unpleasant &     7 &    55 &   \cellcolor{d_small}0.42 &  0.36 \\
Weapon\textsuperscript{$\mathsection$}  &              White &             Black &      Tool &        Weapon &     6 &     7 &                     -1.64 &  1.00 \\
Weapon (Modern)                         &              White &             Black &      Tool &        Weapon &     6 &     9 &                     -1.19 &  0.98 \\
Weight\textsuperscript{\textdagger}     &               Thin &               Fat &  Pleasant &    Unpleasant &    10 &    55 &                     -0.84 &  0.97 \\
\multicolumn{9}{l}{\textsuperscript{$\mathsection$} Originally a picture-IAT (image-only stimuli). \textsuperscript{\textdagger} Originally a mixed-mode IAT (image and verbal stimuli).}\\
\bottomrule
\end{tabular}

\end{table*}

\end{document}